\documentclass[a4paper,10pt]{article}
\usepackage[utf8]{inputenc}
\usepackage{amsthm}
\usepackage{amsfonts}
\usepackage{amssymb}	
\usepackage{amsmath}
\usepackage{bbold}
\allowdisplaybreaks
\usepackage{mathtools}
\usepackage[english]{babel}
\usepackage{color}
\usepackage{slashed}
\usepackage{enumerate}
\usepackage{graphicx}
\usepackage{systeme}
\usepackage{dsfont}
\usepackage{relsize}
\usepackage[margin=0.90in]{geometry}
\usepackage{float}
\usepackage{tikz}
\usepackage[labelformat=simple]{subcaption}

\usepackage{graphicx}
\usepackage{bmpsize}
\usepackage{epstopdf}
\usepackage{bbm}
\usepackage{xcolor,etoolbox}
\usepackage{titling}
\thanksmarkseries{arabic} %command for affiliations with numbers
\usepackage{authblk}
\newcommand*\samethanks[1][\value{footnote}]{\footnotemark[#1]}
\usepackage{blindtext,graphicx}
\usepackage[absolute]{textpos}
\usepackage{physics}
\usepackage[hang,flushmargin]{footmisc}

\usepackage{xfrac}
\usepackage{nicefrac}
\usepackage{esvect}
\usepackage{cases}
\usepackage{empheq}
\usepackage{stmaryrd}
\usepackage{cancel}
\usepackage[linguistics]{forest}
\usepackage{url}
\usepackage[font=small]{caption}
\usepackage{multirow}
\usepackage[giveninits=true,sortcites=true,date=year,maxbibnames=99,doi=false,isbn=false,url=false,eprint=false]{biblatex}
\renewbibmacro{in:}{}
\DeclareFieldFormat{pages}{#1}
\AtEveryBibitem{%
  \clearlist{language}%
}
\usepackage{csquotes}

\usepackage{setspace}
\title{Multi-element metamaterial's design through the relaxed micromorphic model
}
\author{
Leonardo A. Perez Ramirez\thanks{Faculty of Architecture and Civil Engineering, TU Dortmund, August-Schmidt-Str. 8, Dortmund, Germany},
\quad
Gianluca Rizzi\samethanks[1],
\quad and \quad
Angela Madeo\thanks{Head of Chair of Continuum Mechanics, Faculty of Architecture and Civil Engineering, TU Dortmund, \\ \indent \,\,\,\, August-Schmidt-Str. 8, Dortmund, Germany}
}
\begin{document}

\maketitle

Exploring the dynamical response of mechanical metamaterials has gathered increasing attention in the last decades, enabling the design of microstructures exotically interacting with elastic waves (focusing, channeling, band-gaps, negative refraction, cloaking, and many more).
Yet, the application and use of such metamaterials in engineering practice is still deficient due to the lack of effective models unveiling metamaterials’ interactions with more classical materials at finite scales.
In this paper, we show that the relaxed micromorphic model can bring an answer to this open problem and can be effectively used to explore and optimize metamaterials' structures consisting of metamaterials’ and classical materials' bricks of finite size.
We investigate two examples, namely a double-shield structure that can be used to widen the frequency range for which the internal region can be protected and a multiple-shield structure that optimizes both the screening of the regions internal to the single shields and of the zones exterior to the shields themselves.
The exploration of these complex meta-structures has been enabled by the finite element implementation of the relaxed micromorphic model that predicts their response at a fraction of the computational cost when compared to classical simulations.
%%%%%%%%%%%%%%%%%%%
%%%%% SECTION %%%%%
%%%%%%%%%%%%%%%%%%%
\section{Introduction}
\label{sec:rmm_1}
Metamaterials are architectured materials whose mechanical properties go beyond those of classical materials thanks to local resonance or Bragg scattering phenomena occurring in their heterogeneous microstructure.
This micro-heterogeneity allows them to show exceptional mechanical features such as negative Poisson’s ratio \cite{lakes1987foam}, twist and bend in response to being pushed or pulled \cite{frenzel2017three,rizzi2019identification1,rizzi2019identification2}, band-gaps \cite{liu2000locally,wang2014harnessing,bilal2018architected,celli2019bandgap,el2018discrete,koutsianitis2019conventional,goh2019inverse,zhu2015study}, cloaking \cite{buckmann2015mechanical,misseroni2016cymatics,norris2014active,misseroni2019omnidirectional}, focusing \cite{guenneau2007acoustic,cummer2016controlling}, channeling \cite{kaina2017slow,tallarico2017edge,bordiga2019prestress,wang2018channeled,miniaci2019valley}, negative refraction \cite{willis2016negative,bordiga2019prestress,zhu2015study,srivastava2016metamaterial,lustig2019anomalous,morini2019negative}, and many others.

In the last two centuries, the advancement of knowledge on finite-size classical materials modeling has enabled the design of engineering structures (buildings, bridges, airplanes, cars, etc.) resisting to static and dynamic loads.

Today, while the modeling of infinite-size metamaterials is achievable via reliable homogenization techniques, we must acknowledge that such methods are not optimized to deal with finite-size metamaterials' modeling, mainly due to the lack of well-posed homogenized boundary conditions  \cite{chen2001dispersive,willis2009exact,craster2010high,willis2011effective,willis2012construction,boutin2014large,sridhar2018general}.
This conceptual gap has prevented us to explore structures made up of both metamaterials and classical materials, and optimize them towards efficient wave control and energy recovery.

Other enriched models, such as strain-gradient, micropolar, Cosserat or classical micromorphic \cite{dell2012linear,ciallella2022generalized,dell2015elastic,altenbach2010generalized,eremeyev2019two,nazarenko2022variational,eremeyev2019existence,eremeyev2018linear,altenbach2018generalized,eremeyev2017basics,mindlin1963microstructure,eringen1968mechanics} can be used to describe dispersive behaviours or even higher-frequency modes.
However, their use has not been widespread for modeling metamaterials due to limited additional degrees of freedom or to an excessive number of elastic parameters.

In this paper, we show that the mechanical response of finite-size metamaterials can be explored using an elastic- and inertia-augmented micromorphic model (relaxed micromorphic) which is able to describe the main metamaterials’ fingerprint characteristics (anisotropy, dispersion, band-gaps, size-effects, etc.), while keeping a reduced structure (free of unnecessary parameters).
This model can be linked a posteriori to real metamaterials' microstructures via a fitting approach.
The reduced model’s structure, coupled with the introduction of well-posed boundary conditions, allows us to unveil the dynamic response of metamaterials' bricks of finite size and complex shapes.
Playing LEGO with such bricks can enable the design of surprising structures, combining metamaterials and classical materials, actively controlling noise, vibrations, seismic waves, and many others, while being able to recover energy.

At present, the response of finite-size metamaterials’ structures is mostly explored via direct Finite Element (FEM) simulations implementing all microstructures’ details (e.g., \cite{krushynska2017coupling,miniaci2016large,baravelli2013internally,elford2011matryoshka}).
Despite the accurate static and dynamic response that these simulations can provide, they suffer from unsustainable computational costs, already for rather simple structure topoligies.
Therefore, the exploration of large-scale structures combining metamaterials' and classical materials' bricks of different type, size and shape, has been until today out of reach.

In this paper, we lay the basis to overcome this stagnation point and we show how the thorough application of the relaxed micromorphic model can open the way to the design of multi-metamaterial's structures that can control elastic wave propagation.
We will focus here on protection devices (shielding), but similar approach could be adopted for many other applications.
%%%%%%%%%%%%%%%%%%%
%%%% SUBSECTION %%%
%%%%%%%%%%%%%%%%%%%
\section{The relaxed micromorphic model: constitutive laws,\\ equilibrium equations, and boundary conditions}
\label{sec:rmm_2}
In this section, we present the constitutive relations, the equilibrium equations, and the associated boundary conditions for the relaxed micromorphic model \cite{neff2014unifying,voss2022modeling,aivaliotis2020frequency,ghiba2015relaxed,madeo2015wave,neff2015relaxed,neff2020identification}.
The equilibrium equations and the boundary conditions can be derived by means of a variational approach through the associated Lagrangian\footnote{$\langle\cdot , \cdot\rangle$ is the scalar product between tensors of order greater than zero, $\dot{(\cdot)}$ is a derivative with respect to time.}
\begin{equation}
\mathcal{L} \left(\dot{u},\nabla \dot{u}, \dot{P}, \text{Curl} \dot{P}, \nabla u, P, \text{Curl} P\right)
\coloneqq
K \left(\dot{u},\nabla \dot{u}, \dot{P}\right) 
-
W \left(\nabla u, P, \text{Curl} P\right)
\ ,
\end{equation}
where $K$ and $W$ are the kinetic and strain energy, respectively \cite{voss2022modeling}, defined as:
\begin{align}
K \left(\dot{u},\nabla \dot{u}, \dot{P}\right) 
=&
\dfrac{1}{2}\rho \, \langle \dot{u},\dot{u} \rangle + 
\dfrac{1}{2} \langle \mathbb{J}_{\rm m}  \, \text{sym} \, \dot{P}, \text{sym} \, \dot{P} \rangle 
+ \dfrac{1}{2} \langle \mathbb{J}_{\rm c} \, \text{skew} \, \dot{P}, \text{skew} \, \dot{P} \rangle
\notag
\\
&
+ \dfrac{1}{2} \langle \mathbb{T}_{\rm e} \, \text{sym}\nabla \dot{u}, \text{sym}\nabla \dot{u} \rangle
+ \dfrac{1}{2} \langle \mathbb{T}_{\rm c} \, \text{skew}\nabla \dot{u}, \text{skew}\nabla \dot{u} \rangle
\label{eq:kinEneMic}
\\
& 
+ \dfrac{1}{2} \langle \mathbb{M}_{\rm s} \, \text{sym} \, \text{Curl} \, \dot{P}, \text{sym} \, \text{Curl} \, \dot{P} \rangle
% \notag
% \\
% &
+ \dfrac{1}{2} \langle \mathbb{M}_{\rm a} \, \text{skew} \, \text{Curl} \, \dot{P}, \text{skew} \, \text{Curl} \, \dot{P} \rangle
\, ,
\notag
\\
W \left(\nabla u, P, \text{Curl} P\right)
=& 
\dfrac{1}{2} \langle \mathbb{C}_{\rm e} \, \text{sym}\left(\nabla u -  \, P \right), \text{sym}\left(\nabla u -  \, P \right) \rangle
\notag
\\
&
+ \dfrac{1}{2} \langle \mathbb{C}_{\rm c} \, \text{skew}\left(\nabla u -  \, P \right), \text{skew}\left(\nabla u -  \, P \right) \rangle
\notag
\\
&
+ \dfrac{1}{2} \langle \mathbb{C}_{\rm micro} \, \text{sym}  \, P,\text{sym}  \, P \rangle
\notag
\\
& 
+ \dfrac{1}{2} \langle \mathbb{L}_{\rm s} \, \text{sym} \, \text{Curl} P, \text{sym} \, \text{Curl} \, P \rangle
% \notag
% \\
% &
+ \dfrac{1}{2} \langle \mathbb{L}_{\rm a} \, \text{skew} \, \text{Curl} P, \text{skew} \, \text{Curl} \, P \rangle
\, ,
\notag
\end{align}
where $u \in \mathbb{R}^{3}$ is the macroscopic displacement field, $P \in \mathbb{R}^{3\times 3}$ is the non-symmetric micro-distortion tensor, $\rho$ is the macroscopic apparent density, $\mathbb{J}_{\rm m}$, $\mathbb{J}_{\rm c}$, $\mathbb{T}_{\rm e}$, $\mathbb{T}_{\rm c}$, $\mathbb{M}_{\rm s}$, $\mathbb{M}_{\rm a}$, are 4th order micro-inertia tensors, and $\mathbb{C}_{\rm e}$, $\mathbb{C}_{\rm m}$, $\mathbb{C}_{\rm c}$, $\mathbb{L}_{\rm s}$, $\mathbb{L}_{\rm a}$ are 4th order elasticity tensors.
Further details on the definition of these tensors can be found in \cite{rizzi2022boundary,voss2022modeling}.
In the tetragonal symmetry case, these tensors in Voigt notation can be expressed as follows
\begin{align}
        \mathbb{C}_{\rm e}
        &= 
        \begin{pmatrix}
        \lambda_{\rm e} + 2\mu_{\rm e}    & \lambda_{\rm e}                    & \star     & \dots     & 0\\ 
        \lambda_{\rm e}                      & \lambda_{\rm e} + 2\mu_{\rm e}	& \star     & \dots		& 0\\ 
        \star                                   & \star                                 & \star     & \dots     & 0\\
        \vdots                                  & \vdots                                & \vdots    & \ddots	&  \\ 
        0					                    & 0                                     & 0         & 			& \mu_{\rm e}^{*}\\ 
        \end{pmatrix}
        =
        \begin{pmatrix}
        \kappa_{\rm e} + \mu_{\rm e}	& \kappa_{\rm e} - \mu_{\rm e}				& \star & \dots	& 0\\ 
        \kappa_{\rm e} - \mu_{\rm e}	& \kappa_{\rm e} + \mu_{\rm e} & \star & \dots & 0\\
        \star & \star & \star & \dots & 0\\
        \vdots & \vdots	& \vdots & \ddots &\\ 
        0 & 0 & 0 & & \mu_{\rm e}^{*}
        \end{pmatrix},
        \label{eq:micro_ine_1}
        \\[2mm]
        %%%%%%%%%%%%%%%%%%%%%%
        \mathbb{C}_{\rm micro}
        &=
        \begin{pmatrix}
        \lambda_{\rm m} + 2\mu_{\rm m}	        & \lambda_{\rm m}				        & \star     & \dots		& 0\\ 
        \lambda_{\rm m}				            & \lambda_{\rm m} + 2\mu_{\rm m}	    & \star     & \dots		& 0\\ 
        \star                                   & \star                                 & \star     & \dots     & 0\\
        \vdots					                & \vdots					            & \vdots	& \ddots	&  \\ 
        0					                    & 0					                    & 0     	&           & \mu_{\rm m}^{*}\\ 
        \end{pmatrix}
        = 
        \begin{pmatrix}
        \kappa_{\rm m} + \mu_{\rm m}	& \kappa_{\rm m} - \mu_{\rm m}				& \star & \dots	& 0\\ 
        \kappa_{\rm m} - \mu_{\rm m}	& \kappa_{\rm m} + \mu_{\rm m} & \star & \dots & 0\\
        \star & \star & \star & \dots & 0\\
        \vdots & \vdots	& \vdots & \ddots &\\ 
        0 & 0 & 0 & & \mu_{\rm m}^{*}
        \end{pmatrix},
        \notag
        \\[2mm]
        %%%%%%%%%%%%%%%%%%%%%%
        \mathbb{J}_{\rm m}
        &=
        \begin{pmatrix}
        \eta_{3} + 2\eta_{1} & \eta_{3}               & \star       & \dots 			& 0 \\ 
        \eta_{3}             & \eta_{3} + 2\eta_{1}   & \star       & \dots 			& 0 \\ 
        \star                & \star                  & \star       & \dots 			& 0 \\  
        \vdots               & \vdots                 & \vdots      & \ddots 			&   \\ 
        0                    & 0                      & 0       	&                  & \eta^{*}_{1}          \\ 
        \end{pmatrix}
        =
        \rho L_{\rm c}^2
        \begin{pmatrix}
        \kappa_\gamma + \gamma_{1} & \kappa_\gamma - \gamma_{1} & \star & \dots & 0\\ 
        \kappa_\gamma - \gamma_{1} & \kappa_\gamma + \gamma_{1} & \star & \dots & 0\\ 
        \star & \star & \star & \dots & 0\\
        \vdots & \vdots & \vdots & \ddots &\\ 
        0 & 0 & 0 & & \gamma^{*}_{1}\\ 
        \end{pmatrix},
        \notag
\end{align}
	%%%%%%%%%%%%%%%%%%%%%%
\begin{align}
        \mathbb{T}_{\rm e}
        &=
        \begin{pmatrix}
        \overline{\eta}_{3} + 2\overline{\eta}_{1}	& \overline{\eta}_{3}        			   	    & \star		& \dots     & 0 \\ 
        \overline{\eta}_{3}         				& \overline{\eta}_{3} + 2\overline{\eta}_{1} 	& \star		& \dots     & 0 \\ 
        \star                                       & \star                                         & \star     & \dots 	& 0 \\  
        \vdots                    					& \vdots                 			    	   	& \vdots    & \ddots 	&	\\ 
        0                   				    	& 0								            	& 0 		&           &\overline{\eta}^{*}_{1}
        \end{pmatrix}
        =
        \rho L_{\rm c}^2
        \begin{pmatrix}
        \overline{\kappa}_{\gamma} + \overline{\gamma}_{1} & \overline{\kappa}_{\gamma} - \overline{\gamma}_{1} & \star & \dots	& 0\\ 
        \overline{\kappa}_{\gamma} - \overline{\gamma}_{1} &   \overline{\kappa}_{\gamma} + \overline{\gamma}_{1} & \star & \dots & 0\\ 
        \star & \star & \star & \dots & 0\\
        \vdots & \vdots & \vdots & \ddots &\\ 
        0 & 0 & 0 & & \overline{\gamma}^{*}_{1}
        \end{pmatrix},
        \label{eq:micro_ine_2}
        \\[2mm]
        %%%%%%%%%%%%%%%%%%%%%%
        \mathbb{L}_{\rm s}
        &=
        L_{\rm c}^2
        \begin{pmatrix}
        \star                   & \star                   & \star                   & \multicolumn{2}{c}{\dots} & 0     \\
        \star                   & \star                   & \star                   & \multicolumn{2}{c}{\dots} & 0     \\
        \star                   & \star                   & \star                   & \multicolumn{2}{c}{\dots} & 0     \\
        \multirow{2}{*}{\vdots} & \multirow{2}{*}{\vdots} & \multirow{2}{*}{\vdots} & \alpha_1    & 0           & 0     \\
        &                         &                         & 0           & \alpha_1    & 0     \\
        0                       & 0                       & 0                       & 0           & 0           & \star
        \end{pmatrix}
        ,
        %%%%%%%%%%%%%%%%%%%%%%
        \qquad\qquad
        \mathbb{M}_{\rm s}
        =
        \rho L_{\rm c}^4
        \begin{pmatrix}
        \star                   & \star                   & \star                   & \multicolumn{2}{c}{\dots} & 0     \\
        \star                   & \star                   & \star                   & \multicolumn{2}{c}{\dots} & 0     \\
        \star                   & \star                   & \star                   & \multicolumn{2}{c}{\dots} & 0     \\
        \multirow{2}{*}{\vdots} & \multirow{2}{*}{\vdots} & \multirow{2}{*}{\vdots} & \beta_1    & 0           & 0     \\
        &                         &                         & 0           & \beta_1    & 0     \\
        0                       & 0                       & 0                       & 0           & 0           & \star
        \end{pmatrix},
        \notag
        \\[2mm]
        %%%%%%%%%%%%%%%%%%%%%%
        \noalign{\centering
        $\mathbb{J}_{\rm c}
        =
        \begin{pmatrix}
        \star   & 	0		& 0\\ 
        0       & \star 	& 0\\ 
        0       &   0       & 4\eta_{2}
        \end{pmatrix}
        =
        \rho L_{\rm c}^2
        \begin{pmatrix}
        \star & 0 & 0\\ 
        0 & \star & 0\\ 
        0 & 0 & 4\,\gamma_{2}
        \end{pmatrix},\qquad
        %%%%%%%%%%%%%%%%%%%%%%
        \mathbb{T}_{\rm c}
        =
        \begin{pmatrix}
        \star   & 0         & 0\\ 
        0       & \star     & 0\\ 
        0       & 0         & 4\overline{\eta}_{2}
        \end{pmatrix}
        =
        \rho L_{\rm c}^2
        \begin{pmatrix}
        \star & 0 & 0\\ 
        0 & \star & 0\\ 
        0 & 0 & 4\,\overline{\gamma}_{2}
        \end{pmatrix}.$\\[2mm]
        %%%%%%%%%%%%%%%%%%%%%%
        $\mathbb{C}_{\rm c}
        = 
        \begin{pmatrix}
        \star & 0 & 0\\ 
        0 & \star & 0\\ 
        0 & 0 & 4\,\mu_{\rm c}
        \end{pmatrix},
        %%%%%%%%%%%%%%%%%%%%%%
        \qquad
        \mathbb{L}_{\rm a}=
        L_{\rm c}^2
        \begin{pmatrix}
        4\,\alpha_2 & 0 & 0\\ 
        0 & 4\,\alpha_2 & 0\\ 
        0 & 0 & \star
        \end{pmatrix},
        %%%%%%%%%%%%%%%%%%%%%%
        \qquad
        \mathbb{M}_{\rm a}
        =
        \rho \, L_{\rm c}^4
        \begin{pmatrix}
        4\,\beta_2 & 0 & 0\\ 
        0 & 4\,\beta_2 & 0\\ 
        0 & 0 & \star
        \end{pmatrix},$\\
        }\notag
\end{align}
Only the in-plane components are reported since these are the only ones that play a role in the plane-strain simulations presented in the following sections.
%%%%%%%%%%%%%%%%%%%
%%%% SUBSECTION %%%
%%%%%%%%%%%%%%%%%%%
\subsection{Equilibrium equations}
\label{subsec:2}
The action functional $\mathcal{A}$ is defined as
\begin{align}
\mathcal{A}=\iint\limits_{\Omega \times \left[0,T\right]} 
% \hspace{-1em}
\mathcal{L} \left(\dot{u},\nabla \dot{u}, \dot{P}, \text{Curl} \,  \dot{P}, \nabla u, P, \text{Curl} \,  P\right)
\, dx \, dt \, .
\label{eq:action_func}
\end{align}
The first variation $\delta \mathcal{A}$ of the action functional can be related to the virtual work of the internal forces as
\begin{align}
\delta \mathcal{A}
\coloneqq
\int_{0}^{T} \,  \mathcal{W}^{\text{int}} \, dt 
\, .
\label{eq:first_var_A}
\end{align}
Here, the variation operator $\delta$ indicates variation with respect to the kinematic fields $(u,P)$.
Furthermore, it follows from the least-action principle that $\delta \mathcal{A}=0$ uniquely defines both the equilibrium equations and the boundary conditions (both Neumann and Dirichlet).
Thus, the relaxed micromorphic equilibrium equations in strong form are
\begin{equation}
\rho\,\ddot{u} - \text{Div}\widehat{\sigma} = \text{Div}\widetilde{\sigma}
\,,
\qquad\qquad\qquad
\overline{\sigma} + \text{Curl} \, \overline{m} = \widetilde{\sigma} - s -\text{Curl} \, m
\, ,
\label{eq:equiMic}
\end{equation}
where
\begin{align}
\widetilde{\sigma}
&
\coloneqq \mathbb{C}_{\rm e}\,\text{sym}(\nabla u-P) + \mathbb{C}_{\rm c}\,\text{skew}(\nabla u-P)
\, ,
&
\widehat{\sigma}
&
\coloneqq \mathbb{T}_{\rm e}\,\text{sym} \, \nabla\ddot{u} + \mathbb{T}_{\rm c}\,\text{skew} \, \nabla\ddot{u}
\,,
\notag
\\
m
&
\coloneqq
\mathbb{L}_{\rm s}\,\text{sym}\, \text{Curl} \,  P + \mathbb{L}_{\rm a}\,\text{skew}\, \text{Curl} \,  P
\,,
&
\overline{\sigma}
&
\coloneqq \mathbb{J}_{\rm m}\,\text{sym} \, \ddot{P} + \mathbb{J}_{\rm c}\,\text{skew} \, \ddot{P}
\,,
\label{eq:equiSigAll}
\\
\overline{m}
&
\coloneqq \mathbb{M}_{\rm s}\,\text{sym} \, \text{Curl} \, \ddot{P} + \mathbb{M}_{\rm a}\,\text{skew} \, \text{Curl}\ddot{P}
\,,
&
s
&
\coloneqq \mathbb{C}_{\rm micro}\, \text{sym} P
\, .
\notag
\end{align}
The associated boundary conditions are presented in Sect. \ref{sec:rmm_2.2}.

%%%%%%%%%%%%%%%%%%%
%%%% SUBSECTION %%%
%%%%%%%%%%%%%%%%%%%
\subsection{Boundary and interface conditions}
\label{sec:rmm_2.2}
Well-posed boundary conditions are essential for the analysis of macroscopic metamaterial's samples of finite size.
We introduce the work of external surface forces $\mathcal{W}^{\rm ext}$ acting on $\partial \Omega$ over the time interval $\left[0,T\right]$ for the relaxed micromorphic model as
\begin{align}
\int_{0}^{T} \, \mathcal{W}^{\rm ext} \, dt
=
\iint\limits_{\partial\Omega \times \left[0,T\right]}
% \hspace{-1em}
\langle f^{\rm ext} , \delta u \rangle \, dx \, dt
+
\iint\limits_{\partial\Omega \times \left[0,T\right]}
% \hspace{-1em}
\langle \Phi^{\rm ext} , \delta P \rangle \, dx \, dt \, ,
\end{align}
\noindent
where $f^{\rm ext}$ and $\Phi^{\rm ext}$ are the external surface forces and double forces, respectively.

%%%%%% FIGURE %%%%%
%%%%%%%%%%%%%%%%%%%
\begin{figure}[!h]
\centering
\includegraphics[width=0.8\textwidth]{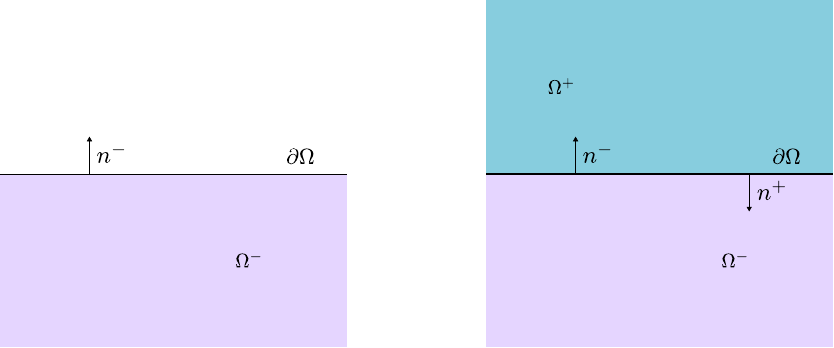}
\caption{Diagram of the boundary and interface of a relaxed micromorphic continuum: (\textit{left}) boundary with free surface, and (\textit{right}) interface between two different relaxed micromorphic continua.
}
\label{fig:interface_conditions}
\end{figure}

According to the Principle of Virtual Work, we can state that the the displacement field $u$ and the micro-distortion field $P$ have to satisfy the equality 
\begin{align}
	\int_{0}^{T}\mathcal{W}^{\text{int}} = \int_{0}^{T} \mathcal{W}^{\rm ext} \, ,
\end{align}
which implies the following duality condition to hold at a free relaxed micromorphic interface
\begin{align}
	\iint\limits_{\partial\Omega \times \left[0,T\right]}
% 	\hspace{-1em}
	\langle t, \delta u \rangle
	=
	\iint\limits_{\partial\Omega \times \left[0,T\right]}
% 	\hspace{-1em}
	\langle f^{\rm ext}, \delta u \rangle \, ,
	\qquad
	\iint\limits_{\partial\Omega \times \left[0,T\right]}
% 	\hspace{-1em}
	\langle \tau, \delta P \rangle
	=
	\iint\limits_{\partial\Omega \times \left[0,T\right]}
% 	\hspace{-1em}
	\langle \Phi^{\rm ext}, \delta P \rangle
	\label{eq:trac_micro_free}
	\, ,
\end{align}
where the generalized traction $t$ and the double traction $\tau$ are defined as
\begin{align}
	t = \left(\widetilde{\sigma} + \widehat{\sigma} \right) \, n \, ,
	\qquad\qquad\qquad
	\tau = m \times n \, ,
	\label{eq:tractions}
\end{align}
while $n$ is the normal at the boundary $\partial \Omega$, and the cross product acts row-wise.
It is also recalled the expression of the traction for a classical Cauchy material
\begin{align}
	t_{\rm Cauchy} = \sigma_{\rm Cauchy} n \, ,
	\label{eq:tractions_Cau}
\end{align}
where $\sigma_{\rm Cauchy}=\lambda \, \text{tr}\left(\text{sym} \nabla u\right) \, \mathbb{1} + 2\mu \, \text{sym}\nabla u$ is the stress tensor for an isotropic linear elastic constitutive model.

The interface conditions between two relaxed micromorphic domains $\Omega^-$ and $\Omega^+$ can be similarly derived (see Fig.~\ref{fig:interface_conditions}) and, in their strong form, they read
\begin{align}
	\begin{cases}
		t^+ = t^- \, ,
		\\
		u^+ = u^- \, ,
	\end{cases}
	\qquad\qquad\qquad
	\begin{cases}
		\tau^+ =\tau^- 
	 \, ,
		\\
		\left(P \times n\right)^+ = \left(P \times n\right)^-
	 \, .
	\end{cases}
	\label{eq:trac_micro_material}
\end{align}

As a particular case, if $\Omega^+$ is a classical Cauchy continuum, then the interface conditions reduce just to
\begin{align}
	\begin{cases}
		t_{\rm Cauchy}^+ = t^- \, ,
		\\
		u^+ = u^- \, ,
	\end{cases}
	\qquad\qquad\quad
	\begin{cases}
	\tau =0 \, ,
	\\
	\text{or}
	\\
	\left(P \times n\right) = 0 \, .
\end{cases}
\hspace{1.2cm}
	\label{eq:trac_micro_material_2}
\end{align}
Since in a classical Cauchy material neither $m$ nor $P$ are defined, from the perspective of the higher order interface conditions the problem reduces to a free-end boundary conditions case.
Therefore we can either chose the Neumann boundary conditions $\tau =0$, which can be interpreted as a ``free microstructure" condition, or the Dirichlet boundary conditions $\left(P \times n\right) = 0$, which can be interpreted as a ``fixed microstructure" condition.

In the reminder of the paper we will set the curvature terms to zero ($\mathbb{L}_{\rm s}=\mathbb{L}_{\rm a}=\mathbb{M}_{\rm s}=\mathbb{M}_{\rm a}=0$ in eq.(\ref{eq:kinEneMic})), since their effect is most relevant in statics than in dynamics \cite{rizzi2022boundary,rizzi2022metamaterial}. It follows that neither the boundary conditions on $\tau$ nor the one on $P$ must be assigned in this particular case.

%%%%%%%%%%%%%%%%%%%
%%%% SUBSECTION %%%
%%%%%%%%%%%%%%%%%%%
\subsection{Numerical results}
\begin{figure}[!h]
	\centering
		\includegraphics[width=0.75\textwidth]{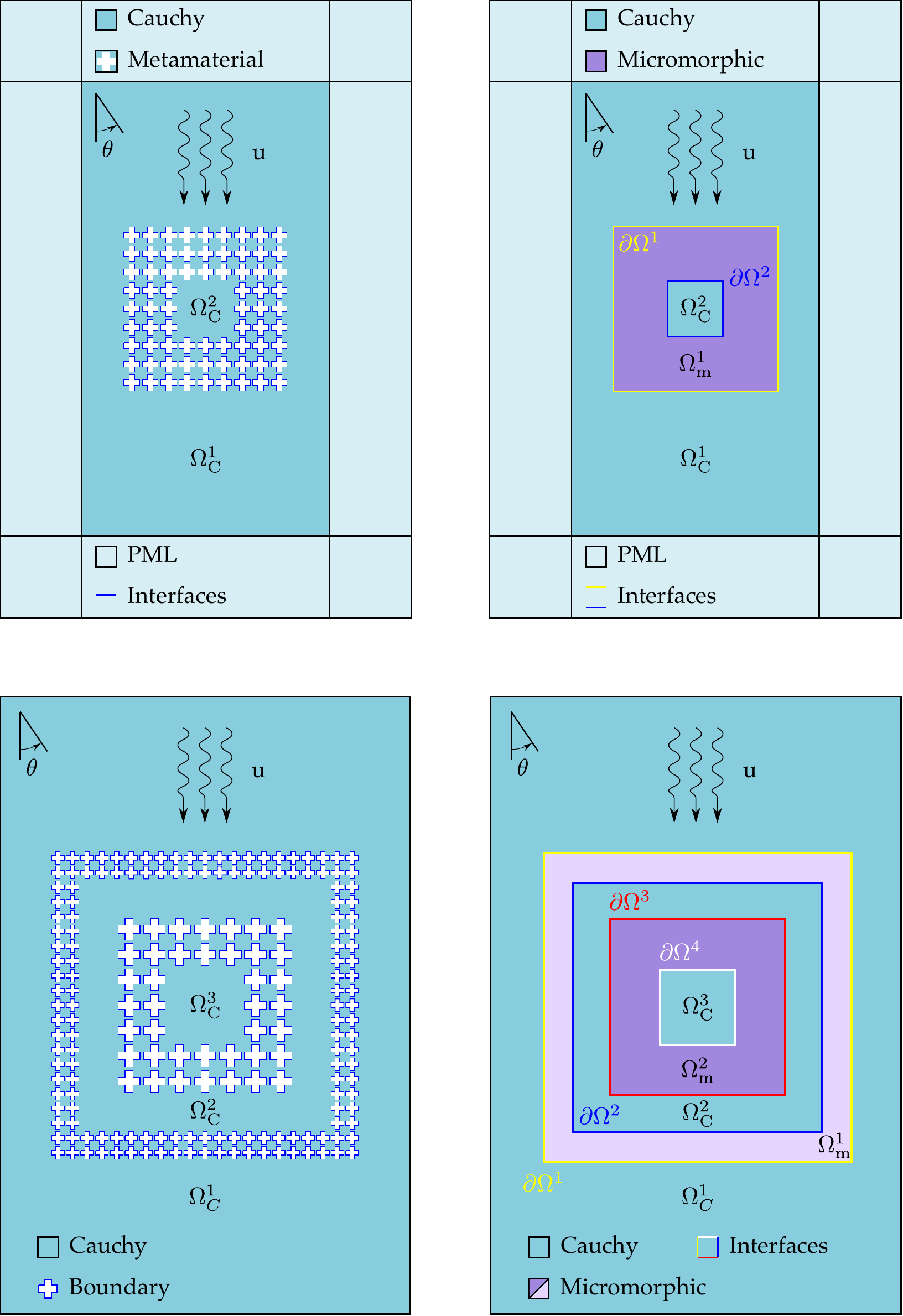}
	\caption{
 Schematics of metamaterials' shields: (\textit{top}) single shield and (\textit{bottom}) double shield. In the latter, the perfectly matched layer surrounding the outermost Cauchy continuum is still present but it is here omitted for better depiction of the details of the double-shield's configuration.
 } 
	\label{fig:domain_single}
\end{figure}

\begin{figure}[!h]
	\begin{minipage}{0.49\textwidth}
		\centering
		\includegraphics[width=0.75\textwidth]{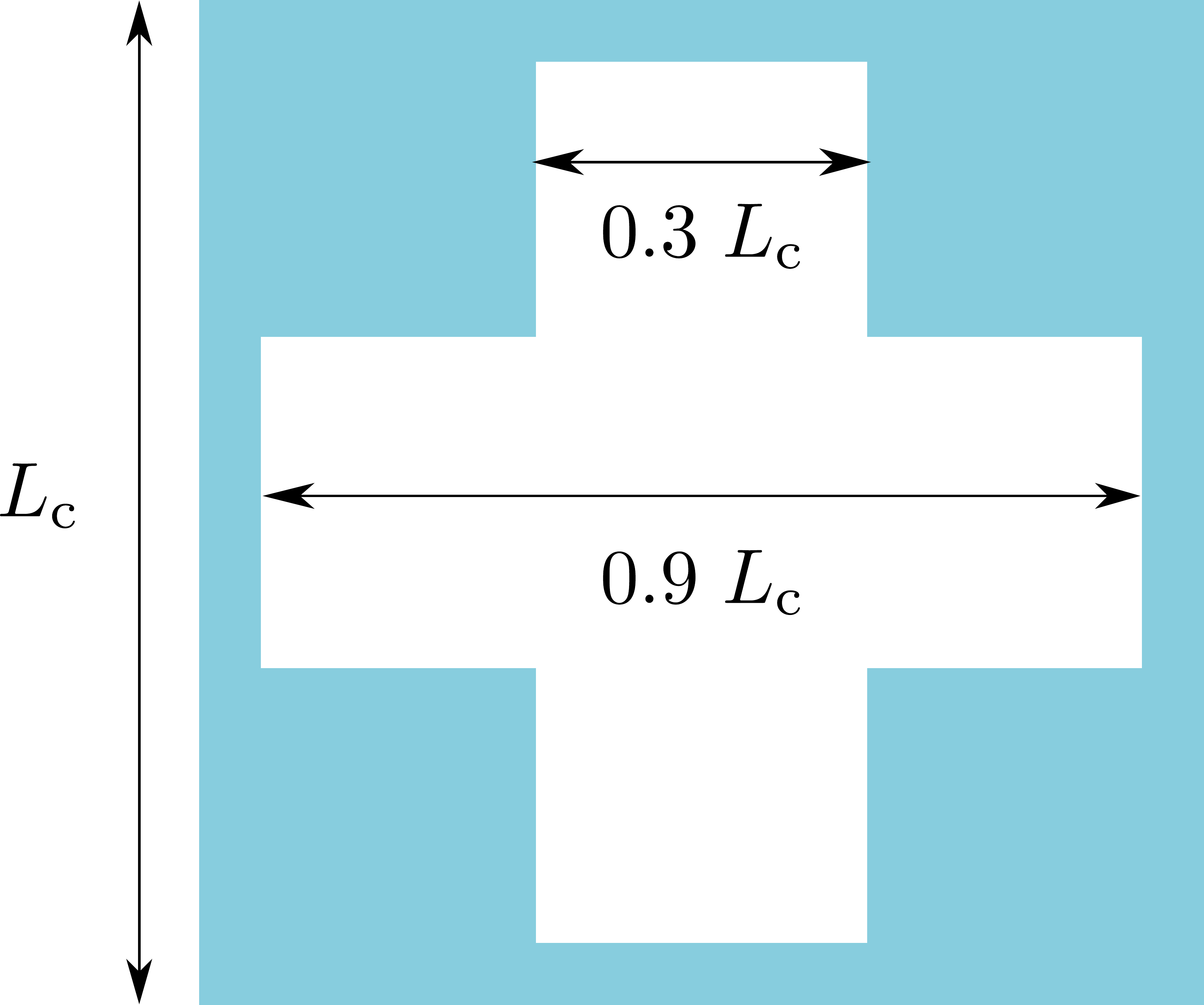}
	\end{minipage}
	\begin{minipage}{0.49\textwidth}
% 			\centering
			\begin{tabular}{cccccc}
				\hline
				$L_{\rm c}$ & $\rho_{\rm Al}$ & $\lambda_{\rm Al}$ & $\mu_{\rm Al}$
				\\[1mm]
				[mm] & [kg/m$^3$] &   [GPa]  &    [GPa]
				\\[1mm]
				\hline
				1 & 2700 & 5.11 & 2.63
				\\
				\hline
				\vspace{0.2cm}
			\end{tabular}
	\end{minipage}
	\caption{(\textit{left panel}) unit cell geometry;
	(\textit{right panel}) material and geometrical properties of the aluminum unit cell: size of the unit cell $L_{\rm c}$, density $\rho_{\rm Al}$, and the Lamé constants $\lambda_{\rm Al}$ and $\mu_{\rm Al}$.}
	\label{fig:unit_cell}
\end{figure}

We perform a time harmonic study to analyze the response of finite size metamaterials using both fully microstructured and relaxed micromorphic models.
An incident plane wave $u^{\rm I}$ travels in the domain $\Omega_{\rm C}^1$ and acts on the single and double shield devices of Fig.~\ref{fig:domain_single}.
The shields are made up of metamaterials, which themselves are made of unit cells of the type shown in Fig.~\ref{fig:unit_cell}.
For the single-shield configuration, the size of the unit cell is $L_{\rm c}=2.5\ \rm mm$, this size corresponds to a scaling factor $s_{\rm F} = 2.5$.
On the double-shield, the metamaterial of the outer shield has $s_{\rm F} = 1$ while the inner shield has $s_{\rm F} = 2.5$.
The properties of the material composing the matrix of the unit cell are reported in Fig.~\ref{fig:unit_cell}. 

\begin{figure}[!h]
        \centering
        \includegraphics[width=\textwidth]{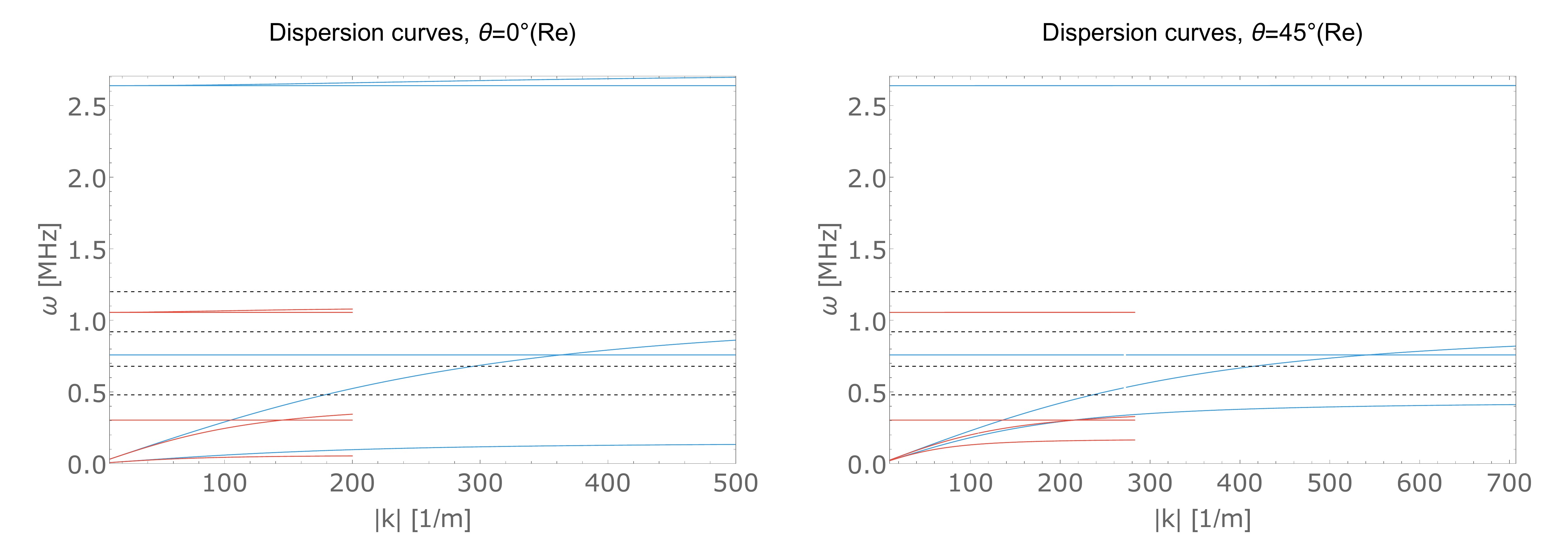}
        \caption{Dispersion curves of the metamaterials issued by the unit cell with $s_{\rm F} = 2.5$ in red and with $s_{\rm F} = 1$ in blue. The dashed lines represent the frequencies $\{0.48,0.68,0.92,1.2\}$ MHz that have been used for the simulations presented in the following sections.
        }
\label{fig:disp_curves}
\end{figure}

The dispersion curves of the two unit cells are presented together in Fig.~\ref{fig:disp_curves}.
It can be seen the widening of the band-gap effect due to the two unit cell's sizes.
The parameters of the relaxed micromorphic model for the unit cell with $s_{\rm F} = 1$ (see also  \cite{rizzi2022boundary}) and $s_{\rm F} = 2.5$ can be found in Table~\ref{tab:parameters_RM}. To obtain the parameters for the unit cell with $s_{\rm F} = 2.5$ form the ones of the unit cell with $s_{\rm F} = 1$, it is just necessary to scale $L_{\rm c}$ with the same factor of the size of the unit cell in eq.(\ref{eq:micro_ine_1})-(\ref{eq:micro_ine_2}) (for more details see \cite{voss2022modeling,demore2022unfolding}).
%%%%%% FIGURE %%%%%
%%%%%%%%%%%%%%%%%%%
%%%%%% TABLE %%%%%%
%%%%%%%%%%%%%%%%%%%
\begin{table}[h!]
    \begin{minipage}{0.49\textwidth}
        \renewcommand{\arraystretch}{1.5}
        \centering
        \resizebox{0.95\textwidth}{!}{%
        \begin{tabular}{ccccccc} 
            $\lambda_{\text{e}}$ [GPa] & $\mu_{\text{e}}$ [GPa] & $\mu^{*}_{\text{e}}$ [GPa]  & $\mu_{\text{c}}$ [GPa]  \\
            \hline\hline
            $2.33$ & $10.92$  & $0.67$ & $2.28\times10^{-3}$ \\
            %\hlinewd{2pt}
            $\lambda_{\rm m}$ [GPa] & $\mu_{\rm m}$ [GPa] & $\mu^{*}_{\rm m}$ [GPa] & $\rho$ [kg/m$^3$]\\
            \hline\hline
            $5.27$  & $12.8$  & $8.33$ & $1485$ \\ 
            %\hlinewd{2pt}
            $\eta_{1}$ [kg/m] & $\eta_{2}$ [kg/m] & $\eta_{3}$ [kg/m] & $\eta^{*}_{1}$ [kg/m]\\
            \hline\hline
            $8.60\times10^{-5}$ & $10^{-7}$ & $-2.20\times10^{-5}$ & $3.30\times10^{-5}$ \\ 
            %\hlinewd{2pt}
            $\overline{\eta}_{1}$ [kg/m] & $\overline{\eta}_{2}$ [kg/m] & $\overline{\eta}_{3}$ [kg/m] & $\overline{\eta}^{*}_{1}$ [kg/m]\\
            \hline\hline
            $5.60\times10^{-5}$  & $7.30\times10^{-4}$ & $1.70\times10^{-4}$ & $9\times10^{-7}$ \\ 
        \end{tabular}%
        }
    \end{minipage}
    \begin{minipage}{0.49\textwidth}
        \renewcommand{\arraystretch}{1.5}
        \centering
        \resizebox{0.95\textwidth}{!}{%
        \begin{tabular}{ccccccc} 
            $\lambda_{\text{e}}$ [GPa] & $\mu_{\text{e}}$ [GPa] & $\mu^{*}_{\text{e}}$ [GPa]  & $\mu_{\text{c}}$ [GPa]  \\
            \hline\hline
            $2.33$ & $10.92$  & $0.67$ & $2.28\times10^{-3}$ \\
            %\hlinewd{2pt}
            $\lambda_{\rm m}$ [GPa] & $\mu_{\rm m}$ [GPa] & $\mu^{*}_{\rm m}$ [GPa] & $\rho$ [kg/m$^3$]\\
            \hline\hline
            $5.27$  & $12.8$  & $8.33$ & $1485$ \\ 
            %\hlinewd{2pt}
            $\eta_{1}$ [kg/m] & $\eta_{2}$ [kg/m] & $\eta_{3}$ [kg/m] & $\eta^{*}_{1}$ [kg/m]\\
            \hline\hline
            $5.38\times10^{-4}$ & $6.25\times10^{-7}$ & $-1.38\times10^{-4}$ & $2.06\times10^{-4}$ \\ 
            %\hlinewd{2pt}
            $\overline{\eta}_{1}$ [kg/m] & $\overline{\eta}_{2}$ [kg/m] & $\overline{\eta}_{3}$ [kg/m] & $\overline{\eta}^{*}_{1}$ [kg/m]\\
            \hline\hline
            $3.50\times10^{-4}$  & $4.56\times10^{-3}$ & $1.06\times10^{-3}$ & $5.63\times10^{-6}$ \\
        \end{tabular}%
        }
    \end{minipage}
    \caption{
    Parameters of the relaxed micromorphic model for the unit cell of Fig.~\ref{fig:unit_cell} when (\textit{left}) $L_{\rm c} = 1$ mm and (\textit{right}) $L_{\rm c} = 2.5$ mm.
    It is highlighted that in order to obtain the parameters for the unit cell that is just homothetically scaled by a factor $s_{\rm F}=2.5$ it is enough to change the value of $L_{\rm c}$ form 1 to 2.5 mm in the expressions reported in equation (\ref{eq:micro_ine_1})-(\ref{eq:micro_ine_2}) (see also \cite{voss2022modeling,demore2022unfolding})
    }
    \label{tab:parameters_RM}
\end{table}
%%%%%% FIGURE %%%%%
%%%%%%%%%%%%%%%%%%%
%%%%%% FIGURE %%%%%
%%%%%%%%%%%%%%%%%%%

All the numerical studies are performed under the plane strain hypothesis.
The incident wave $u^{\rm I}$ travels in the external Cauchy continuum and can be defined either as a pressure $P$ or a shear $S$ wave as
\begin{align}
	u^{\rm I}
	=
	a\,
	\psi^{\rm P/S,I} \, e^{i \left(k_1^{\rm P/S,I} \, x_1 + k_2^{\rm P/S,I} \, x_2 - \omega \, t \right)} \, ,
	\quad
	\text{with}
	\quad
	\left\{
	\begin{array}{ll}
		k_1^{\rm P,I}
		=
		\sqrt{\dfrac{\rho_{\rm Al}}{\lambda_{\rm Al} + 2\mu_{\rm Al}}\omega^2-\left( k_2^{\rm P,I} \right)^2}
		\\*[5mm]
		k_1^{\rm S,I}
		=
		\sqrt{\dfrac{\rho_{\rm Al}}{\mu_{\rm Al}}\omega^2-\left( k_2^{\rm S,I} \right)^2}
	\end{array}
	\right.
	\, ,
	\label{eq:plane_wave_plus_k}
\end{align}
where $a$ is the amplitude, $\psi^{\rm P/S,I}$ is the normalized eigenvector associated with the pressure or shear wave, $k_1^{\rm P/S,I}$ and $k_2^{\rm P/S,I}$ are the components of the wave vector, and $\omega$ is the frequency.

The perfect contact condition at each interface between the relaxed micromorphic continuum and the classical Cauchy continuum (see Fig.~\ref{fig:domain_single}) are reported in eq.(\ref{eq:trac_micro_material_2}).
For the unit cell considered in this work, in the dynamic range studied, the curvature terms have a negligible effect in the metamaterial's response \cite{d2020effective,rizzi2022boundary}.
This allows for setting $\mathbb{L}_{\rm s}=\mathbb{L}_{\rm a}=\mathbb{M}_{\rm s}=\mathbb{M}_{\rm a}=0$, thus the curvature terms vanish and no double traction has to be defined on the boundaries and the interface conditions reduce to just  eq.(\ref{eq:trac_micro_material_2})$_1$.

Let us consider the equilibrium conditions derived from eq.(\ref{eq:trac_micro_material_2}) as they act, for example, on the boundary $\partial \Omega^1$.
The superposition principle allows for the decomposition of the displacement and traction on the Cauchy side, into an incident and scattered component.
The scattered components are indicated by the subscript ``sca'', and the superscripts refer to their domain. The traction $t^{\rm I}$ is that produced by the incident wave and the conditions can be summarized as
\begin{align}
	\left\{
	\arraycolsep=1pt\def\arraystretch{1}
	\begin{array}{ccccccccccccc}
		u^{-} 
		& = 
		& u^{+}
		& \quad \Longleftrightarrow
		& \quad u^{\Omega_{\rm m}^1}
		& = 
		& u^{\Omega_{\rm C}^1}_{\rm sca} + u^{\rm I}
		\\*[2mm]
		t^{-} 
		& = 
		& t^{+}
		& \quad \Longleftrightarrow
		& \quad t^{\Omega_{\rm \rm m}^1}
		& = 
		& t^{\Omega_{\rm C}^1}_{\rm sca} + t^{\rm I}
	\end{array}
	\right.
	\quad \mathrm{on} 
	\quad \partial \Omega^{1} \, .
	\label{eq:jump_cond}
\end{align}
As a consequence, on the boundary $\partial \Omega_{\rm C}^1$, we can then prescribe 
\begin{equation}
	u^{\Omega_{\rm C}^1}_{\mathrm{sca}}
	= 
	u^{\Omega_{\rm m}^1} - u^{\rm I}
	\, ,
	\label{eq:displ_jump_cond}
\end{equation}
and the associated tractions $t^{\Omega_{\rm C}^1}_{\mathrm{sca}}$ and $t^{\Omega_{\rm m}^1}$ derive from the solution of the correspondent displacement fields.
Furthermore, on the boundary, the tractions of $\Omega_{\rm C}^1$ differ with respect to the tractions $\Omega_{\rm m}^1$ by
\begin{equation}
	t^{\Omega_{\rm C}^1}_{\mathrm{sca}} - t^{\Omega_{\rm m}^1}
	= 
	- t^{\rm I}
	\ .
	\label{eq:tract_jump_cond}
\end{equation}
Thus we must prescribe $- t^{\rm I}$ on the boundary $\partial \Omega^1$ of $\Omega_{\rm C}^1$ to ensure the continuity of tractions reported in eq.(\ref{eq:jump_cond}).
For the microstructured material, it is sufficient to prescribe $- t^{\rm I}$ on the free boundary of the holes.
Analogous conditions apply to the other boundaries.
%%%%%%%%%%%%%%%%%%%
%%%%% SECTION %%%%%
%%%%%%%%%%%%%%%%%%%
\section{Parametric study on the thickness of a shielding device: capability limit for the relaxed micromorphic model}
\label{sec:rmm_3}
\begin{figure}[!h]
\centering
\includegraphics[scale=0.64]{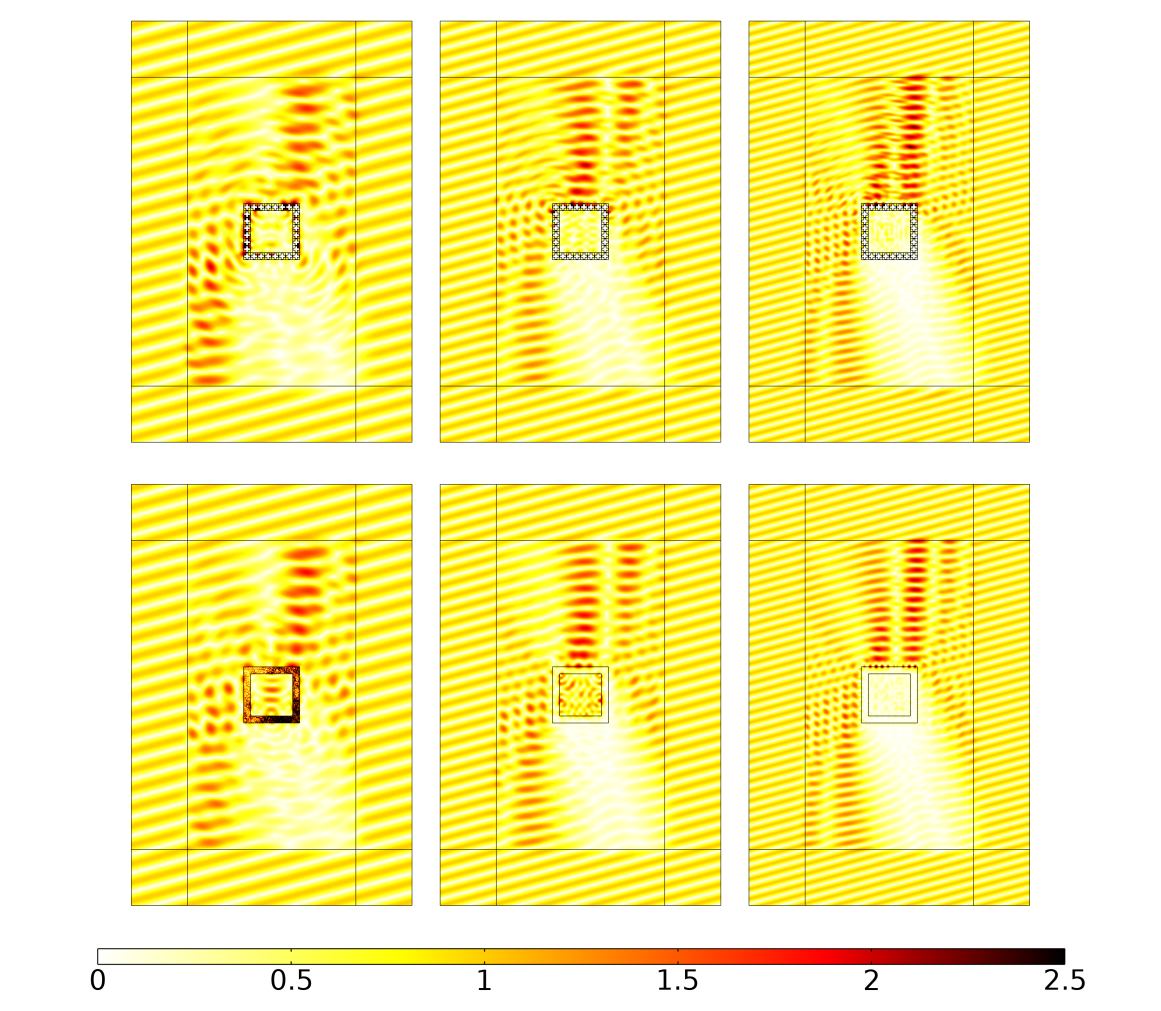}
\caption{Real part of the normalized displacement field of the microstructured material and its equivalent micromorphic continuum for a shield of 1 unit cell thick, for an incident angle $\phi=15^{\circ}$:
(\textit{first row}) microstructured material and 
(\textit{second row}) equivalent micromorphic model for $\omega=0.48$ MHz, $\omega=0.68$ MHz, and $\omega=0.92$ MHz.
All frequencies fall in the band-gap range of the considered metamaterial.}
\label{fig:single_shield_comparison_1}
\end{figure}

In this section, we present the results of a simple shield device modeled both with a full-microstructure model and with the relaxed micromorphic model.
The displacement in the subsequent figures is normalized by the amplitude of the incident wave.
In Fig.~\ref{fig:single_shield_comparison_1}, we can see that a shield device made up of only one unit cell is not able to ensure the desired shielding effect, and this is true both for the microstructured and the relaxed micromorphic models.

\begin{figure}[!h]
\centering
\includegraphics[scale=0.64]{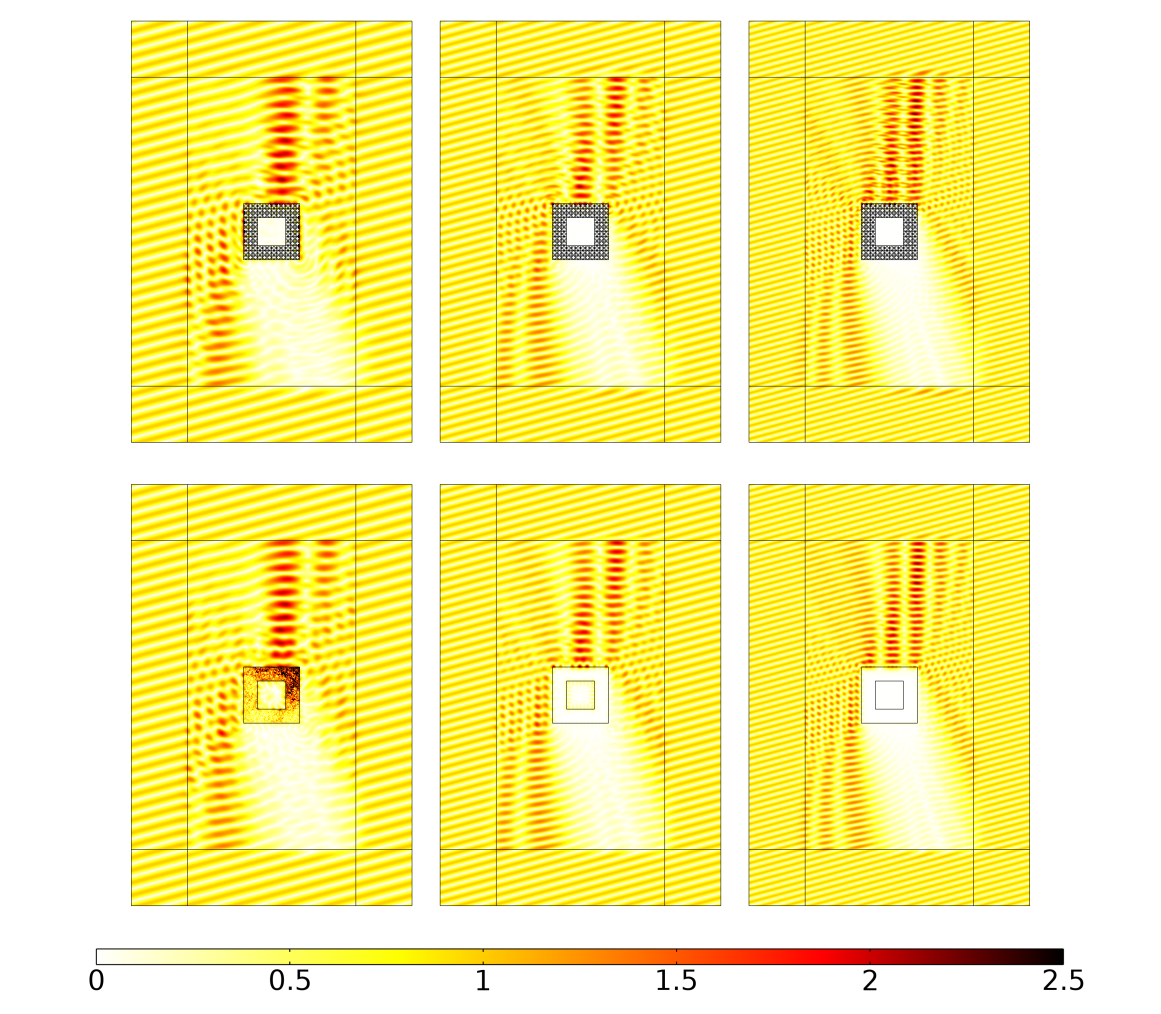}
\caption{Real part of the normalized displacement field of the microstructured material and its equivalent micromorphic continuum for a shield of 3 unit cells thick, for an incident angle $\phi = 15^{\circ}$:
(\textit{first row}) microstructured material and 
(\textit{second row}) equivalent micromorphic model for $\omega=0.48$~MHz, $\omega=0.68$~MHz, and $\omega=0.92$~MHz. 
All frequencies fall in the band-gap region of the considered metamaterial.} 
\label{fig:single_shield_comparison_2}
\end{figure}

\begin{figure}[!h]
\centering
\includegraphics[scale=0.64]{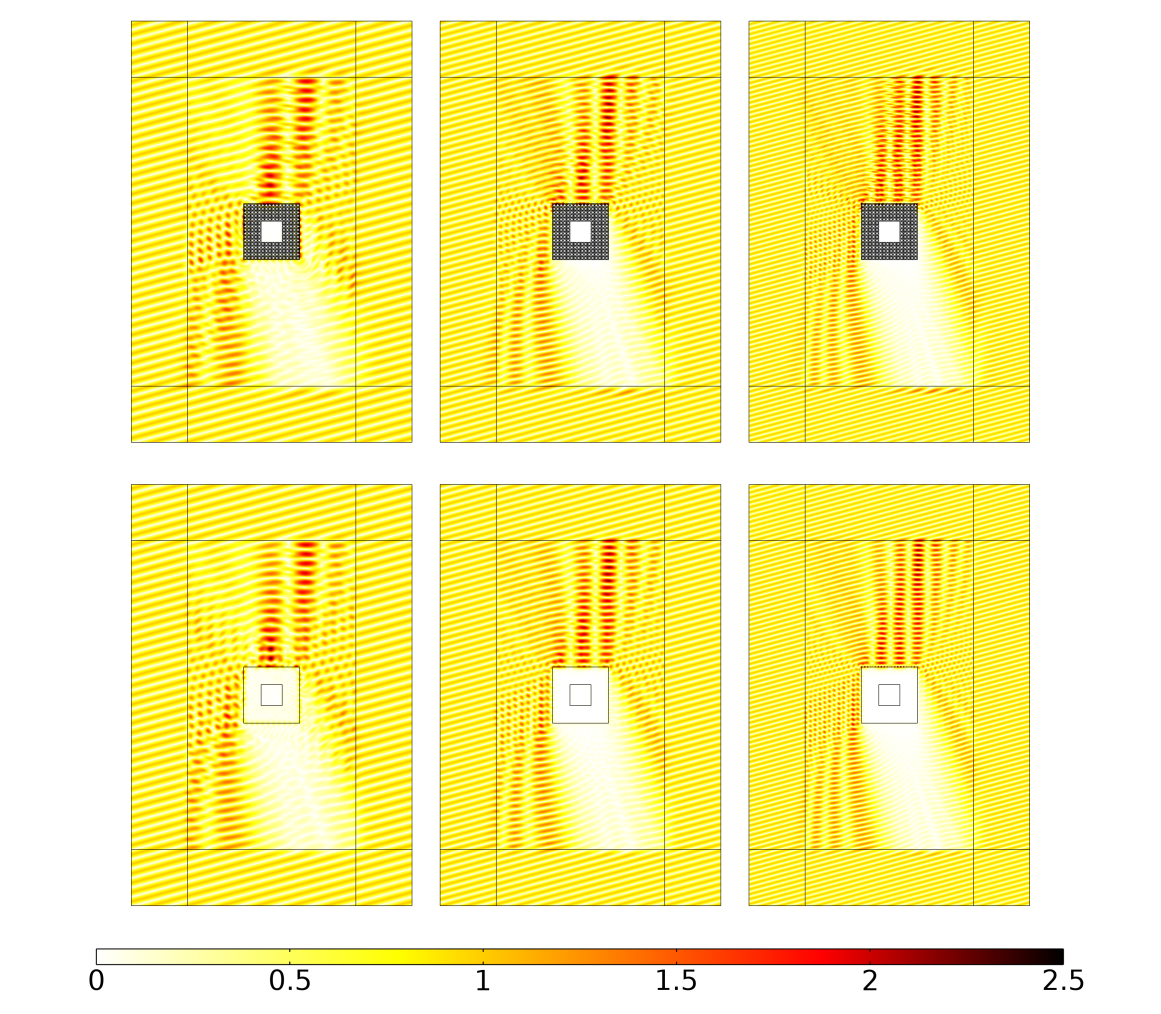}
\caption{Real part of the normalized displacement field of the microstructured material and its equivalent micromorphic continuum for a shield of 5 unit cells thick, for an incident angle $\phi = 15^{\circ}$:
(\textit{first row}) microstructured material and 
(\textit{second row}) equivalent micromorphic model for $\omega=0.48$ MHz, $\omega=0.68$ MHz, and $\omega=0.92$ MHz. 
All frequencies fall in the band-gap region of the considered metamaterial.} 
\label{fig:single_shield_comparison_3}
\end{figure}

\begin{figure}[!h]
\centering
\includegraphics[scale=0.64]{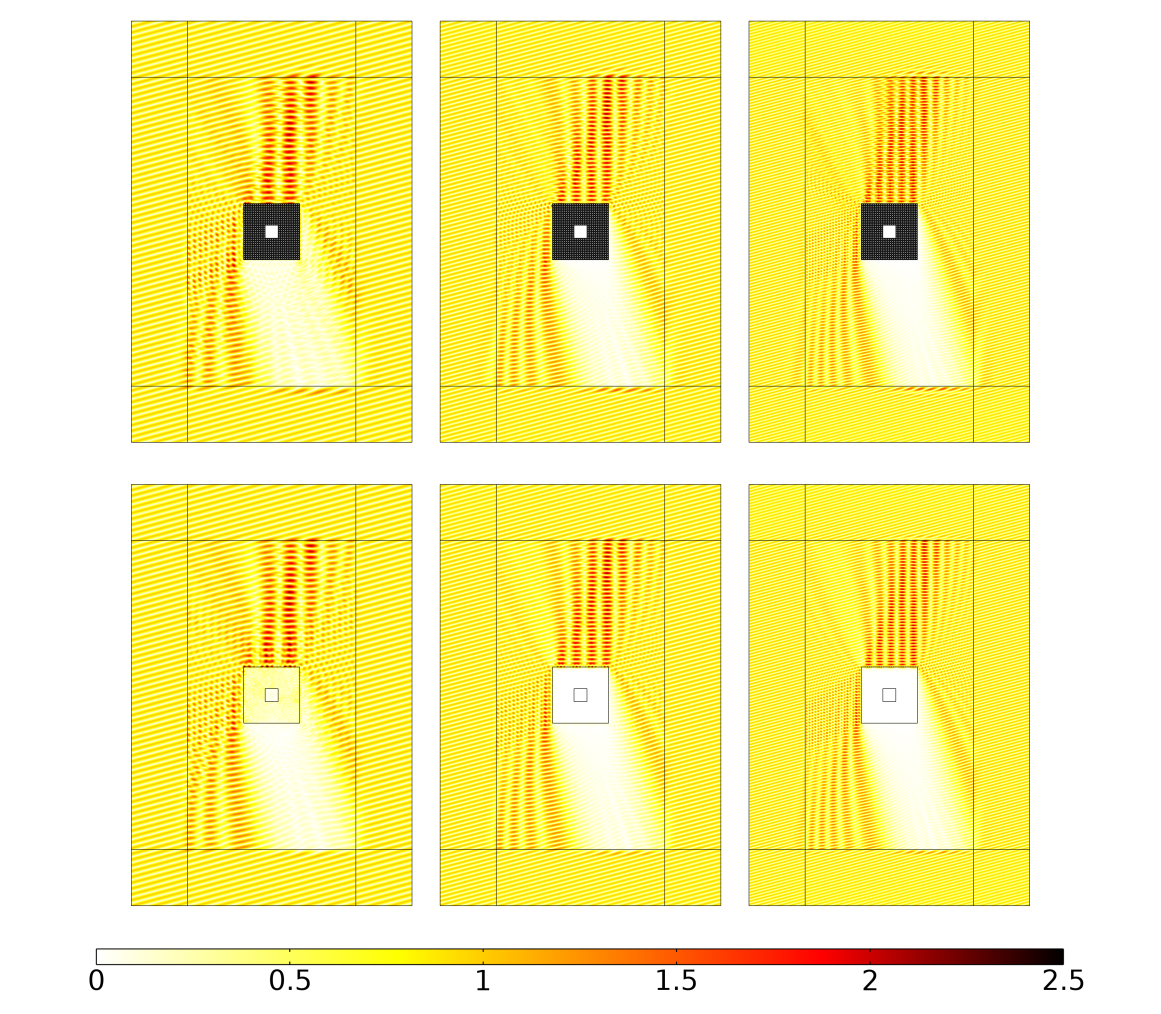}
\caption{Real part of the normalized displacement field of the microstructured material and its equivalent micromorphic continuum for a shield of 10 unit cells thick, for an incident angle $\phi = 15^{\circ}$:
(\textit{first row}) microstructured material and 
(\textit{second row}) equivalent micromorphic model for $\omega=0.48$ MHz, $\omega=0.68$ MHz, and $\omega=0.92$ MHz. 
All frequencies fall in the band-gap region of the considered metamaterial.}
\label{fig:single_shield_comparison_4}
\end{figure}

\begin{figure}[!h]
\centering
\includegraphics[scale=0.64]{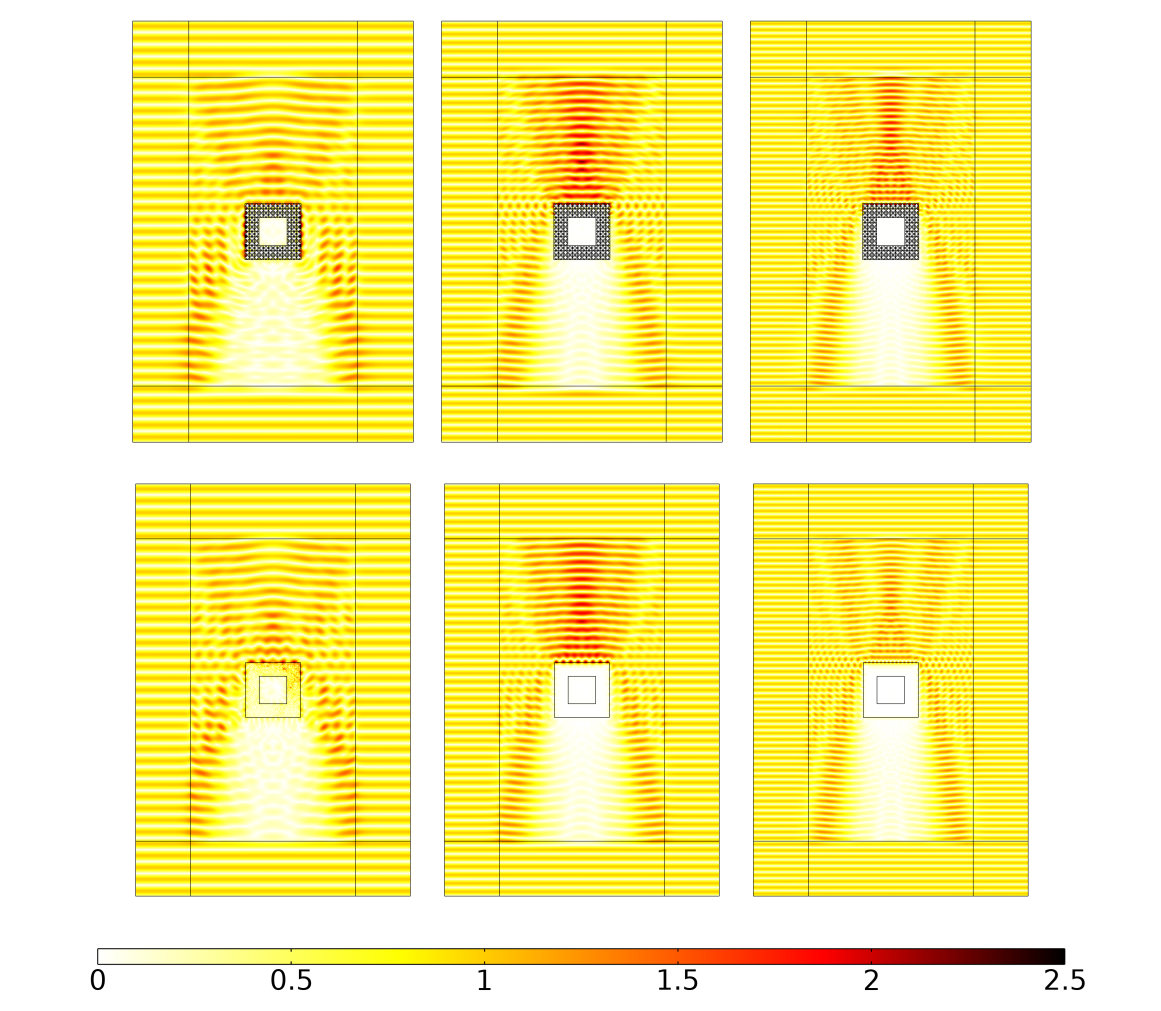}
\caption{Real part of the normalized displacement field of the microstructured material and its equivalent micromorphic continuum for a shield of 3 unit cells thick, for an incident angle $\phi = 0^{\circ}$:
(\textit{first row}) microstructured material and 
(\textit{second row}) equivalent micromorphic model for $\omega=0.48$ MHz, $\omega=0.68$ MHz, and $\omega=0.92$ MHz. 
All frequencies fall in the band-gap region of the considered metamaterial.} 
\label{fig:single_shield_comparison_5}
\end{figure}

We remark that the reflection pattern is nevertheless well captured by the relaxed micromorphic model, even if the shield's thickness is of only one unit cell, while the displacement pattern inside the shield shows a loss of accuracy.
This is due to the fact that the relaxed micromorphic model, being a homogenized model, needs a certain domain size to guarantee results with a high level of accuracy, and we show this by performing a parametric study on the shield's thickness (see Fig.~\ref{fig:single_shield_comparison_1}-\ref{fig:single_shield_comparison_5}).

We can infer from Fig.~\ref{fig:single_shield_comparison_2}-\ref{fig:single_shield_comparison_5} that both the outer and inner displacement fields are well reproduced by the relaxed micromorphic model, as early as the shield is thick enough.
Indeed, we established that both the shielding effect and the relaxed micromorphic model accuracy can be considered to be satisfactory already with a 3 unit cell thickness, especially when considering frequencies in the band-gap range.

All the simulations performed in this section and in the following ones have been run on 2$\times$\textit{AMD EPYC 7453 28-Core Processor}. 
We observe faster computing times for the relaxed micromorphic model than for the full-microstructure simulations, more details are presented in the following sections.

For the sake of completeness, we also present in Fig.~\ref{fig:single_shield_comparison_4} the reflection pattern of a simple shield device which is 3 unit cells thick and for a 0$^{\circ}$ angle of incidence.
Analogous results can be obtained for all other angles of incidence, showing the effectiveness of the relaxed micromorphic model to unveil the macroscopic metamaterial's anisotropy.

%%%%%%%%%%%%%%%%%%%
%%%%% SECTION %%%%%
%%%%%%%%%%%%%%%%%%%
\section{Design of a double shield device}
\label{sec:rmm_4}
In this section, we start exploring multi-shield devices with the aim of increasing the frequency range for which the inner core of the structure is protected from an external excitation.
To do so, we use two different sizes of the unit cell presented in Fig.~\ref{fig:unit_cell} with ${\rm s_F} = 1$ for the outer and ${\rm s_F} = 2.5$ for the inner shield.
In this way the inner structure's core will be protected in the overlapping intervals [1.11,2.65] MHz and [0.44,1.06] MHz.

\begin{figure}[!h]
\centering
\includegraphics[scale=0.64]{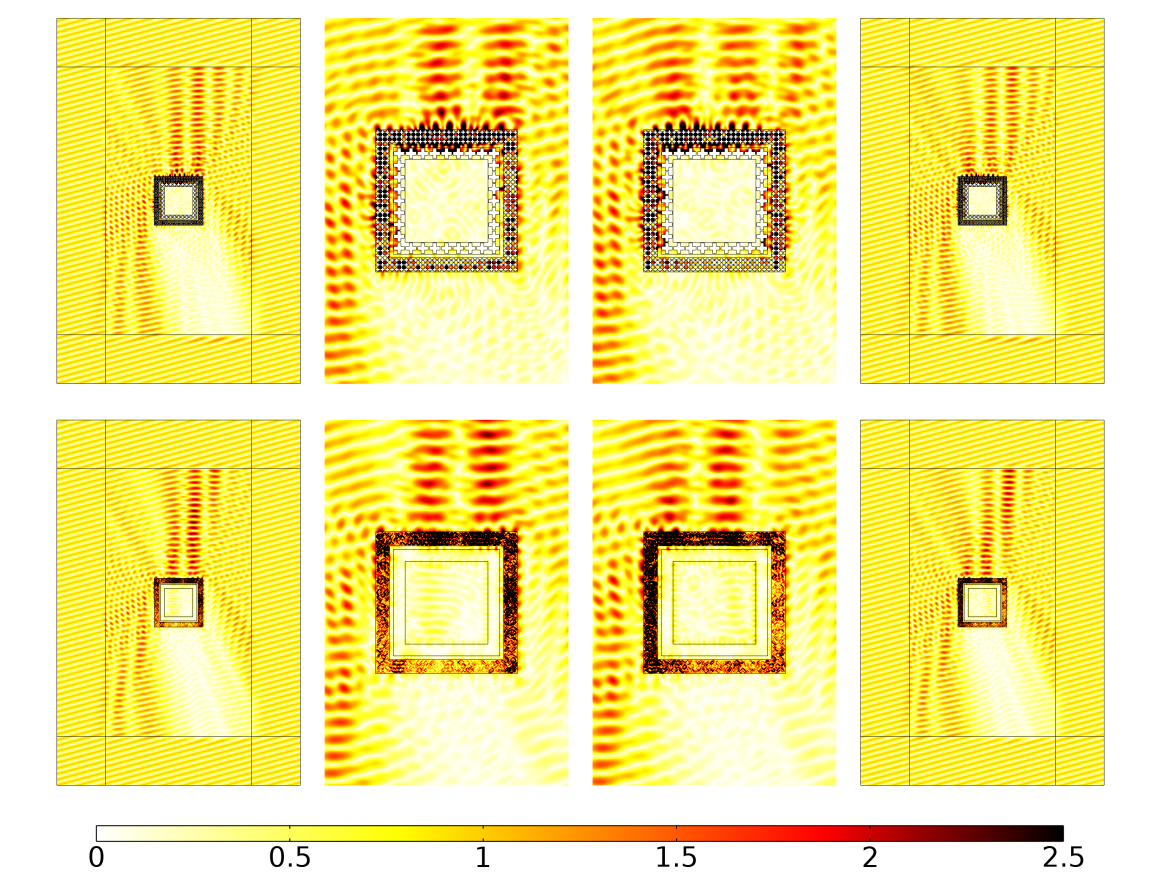}
\caption{Imaginary (\textit{first two columns}) and real (\textit{last two columns}) part of the normalized displacement field of the microstructured material and its equivalent micromorphic continuum:
(\textit{first row}) microstructured material and 
(\textit{second row}) equivalent micromorphic model for $\omega=0.92$ MHz. 
The thickness of the outer shield is 3 unit cells, and the thickness of the inner shield is 1 unit cell.}
\label{fig:double_shield_comparison_1}
\end{figure}

We show in Fig.~\ref{fig:double_shield_comparison_1} a first multi-shield device composed of 3 unit cells for the outer shield and 1 unit cell for the inner shield: the scattering behavior is given for an incidence angle of $15^{\circ}$ and for a frequency of $\omega = 0.92$ MHz, which falls in the band-gap of the outer shield.
We see that the relaxed micromorphic model recovers well the scattering pattern of the double-shield device at a fraction of the computational cost and with lower memory use.
The simulation that uses the relaxed micromorphic model takes 61 s in contrast to the 156 s needed by the geometrically detailed simulation.
The gain in computational time, which is already visible for these simple structures, will be more significant when studying larger structures, as those presented in Sect.~\ref{sec:rmm_5}.

\begin{figure}[!h]
\centering
\includegraphics[scale=0.64]{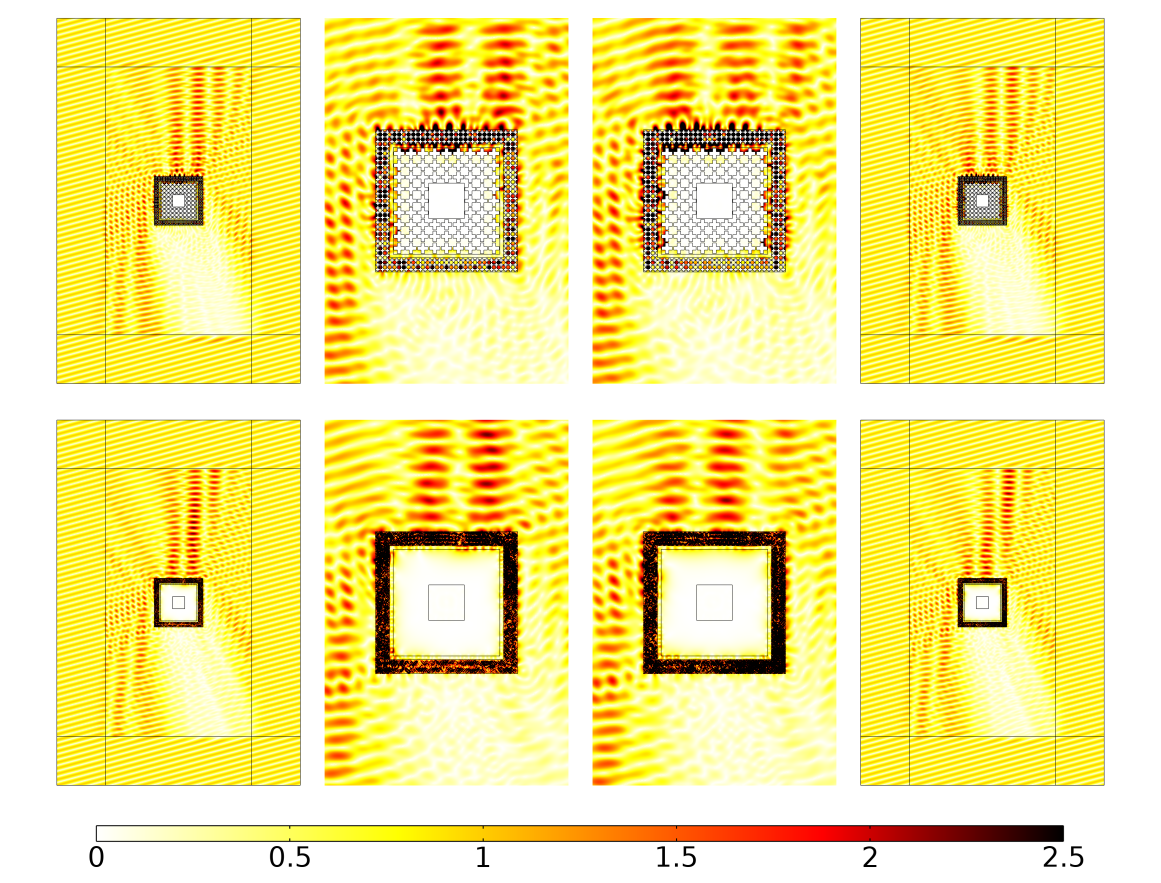}
\caption{Imaginary (first two columns) and real (last two columns) part of the normalized displacement field of the microstructured material and its equivalent micromorphic continuum: 
(\textit{first row}) microstructured material and 
(\textit{second row}) equivalent micromorphic model for $\omega=0.92$ MHz. 
The thickness of the outer shield is 3 unit cells, and the thickness of the inner shield is 3 unit cells.}
\label{fig:double_shield_comparison_2}
\end{figure}

In Fig.~\ref{fig:double_shield_comparison_2} we show the same results for a multi-shield device made of two 3x3 cell shields. The results are again convincing, at a fraction of the computational cost.

\begin{figure}[!h]
\centering
\includegraphics[scale=0.64]{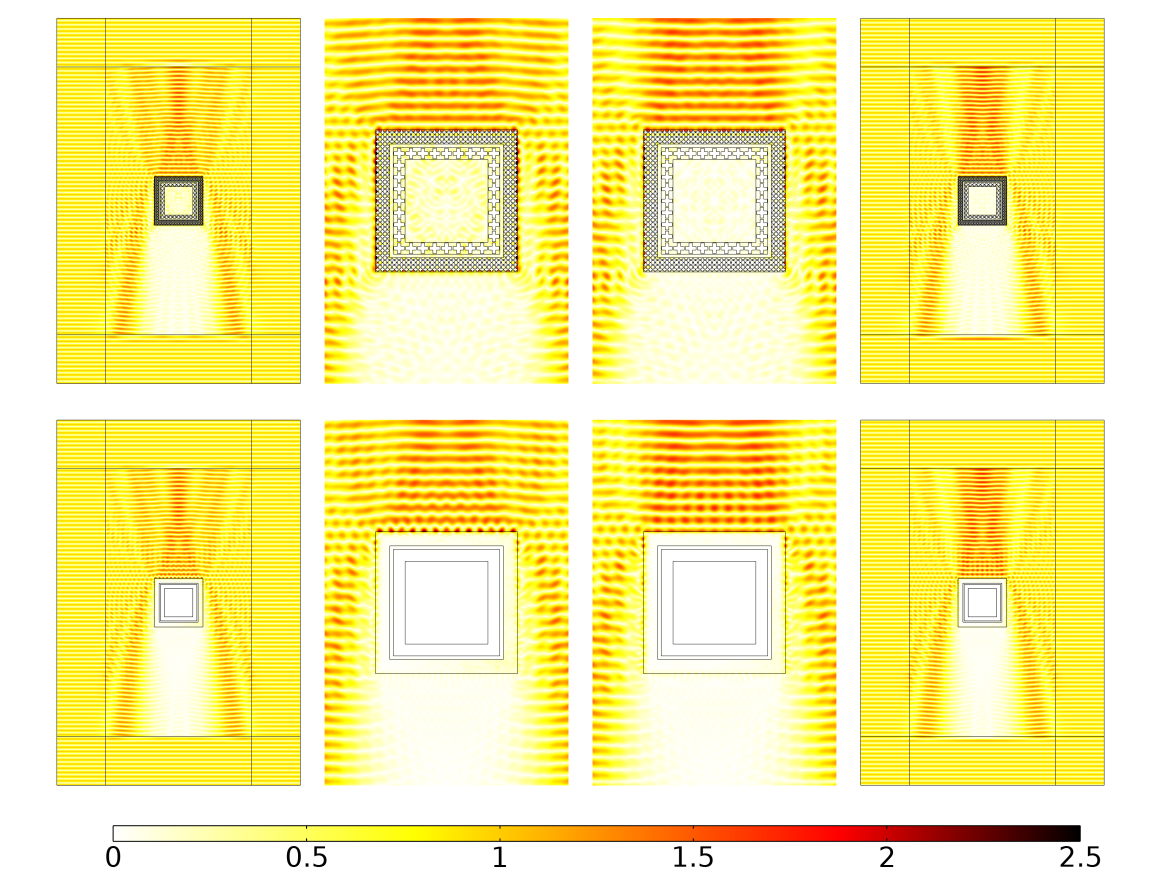}
\caption{Imaginary (first two columns) and real (last two columns) part of the normalized displacement field of the microstructured material and its equivalent micromorphic continuum: 
(\textit{first row}) microstructured material and 
(\textit{second row}) equivalent micromorphic model for $\omega=1.2$ MHz.
The thickness of the outer shield is 3 unit cells, and the thickness of the inner shield is 1 unit cell.}
\label{fig:double_shield_comparison_3}
\end{figure}

Similar results can be shown for other angles of incidence. 
In order to complete our exploration, we now choose a frequency $\omega=1.2$ MHz which falls in the band-gap of the inner shield metamaterial.
Fig.~\ref{fig:double_shield_comparison_3} shows the structure's response for $\omega = 1.2$ MHz and an angle of incidence of $0^{\circ}$. 
Once again, the relaxed micromorphic model performs really well at a fraction of the computational cost when reproducing the displacement field of the outer Cauchy domain.
In this case, the solution of the relaxed micromorphic model takes 94 s when the full-microstructure simulation needs 178 s.

However, these results also showcase a limitation of the micromorphic type models when the curves above the band gap grow flat, as is the case for the unit cell used in this work, see Fig.~\ref{fig:disp_curves}.
In Fig.~\ref{fig:double_shield_comparison_3}, the inner shield does not properly capture the excitation modes at $\omega=1.2$ MHz, and overall there is a slight inaccuracy in the prediction of the interior displacement field as opposed to the displacement present in the full-microstructure simulation. More particularly, the relaxed micromorphic model predicts a complete screening while the microstructured simulations show a small displacement occurring in the interior part. This difference is due, to a big extent, to the fact that we are considering only the first 6 modes for the relaxed micromorphic model, while at the considered frequency higher order modes can be slightly activated.

%%%%%%%%%%%%%%%%%%%
%%%%% SECTION %%%%%
%%%%%%%%%%%%%%%%%%%
\section{Multiple-shields Optimization}
\label{sec:rmm_5}
While metamaterials' modeling has gathered a wealth of research effort in the last decades, the study of large-scale metamaterial's structures consisting of many metamaterials' components has received little attention.
For example, many researchers have successfully proposed metamaterials' shielding configurations \cite{achaoui2017clamped,miniaci2021hierarchical,miniaci2016large,palermo2016engineered}, but no effort is paid to the study of the global reflection pattern of a system consisting of multiple shielding devices.
This lack is mainly due to the computational limitations of full-microstructure simulations, on the one hand, and to lack of knowledge of homogenized approaches on well-posed boundary conditions, on the other.

\begin{figure}[!h]
\centering
\includegraphics[scale=0.64]{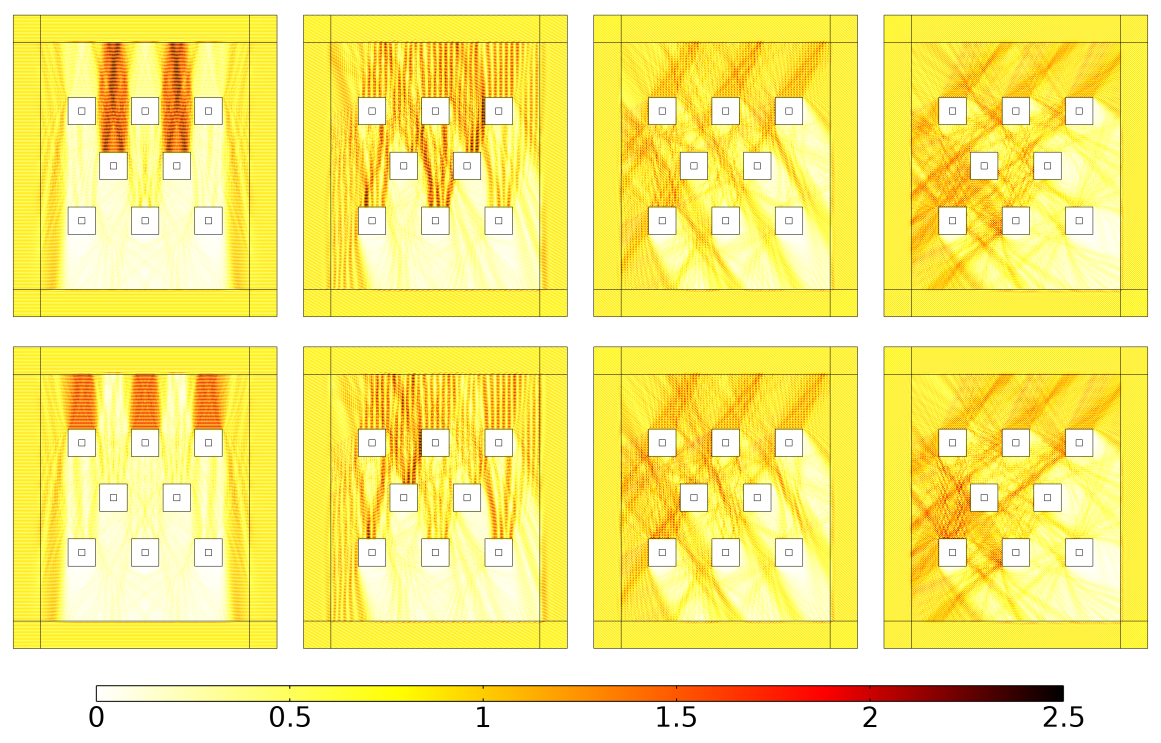}
\caption{Normalized displacement field of a multiple shield structure for an incident wave of frequency $\omega=0.92$ MHz: 
(\textit{first row}) imaginary and 
(\textit{second row}) real part of the displacement. 
Form \textit{left} to \textit{right} the angle of the incident wave is $\phi=0^{\circ}$, $\phi=15^{\circ}$, $\phi=30^{\circ}$, and $\phi=45^{\circ}$. The thickness of each single shields is 10 unit cells.}
\label{fig:multiple-3x3}
\end{figure}

\begin{figure}[!h]
\centering
\includegraphics[scale=0.64]{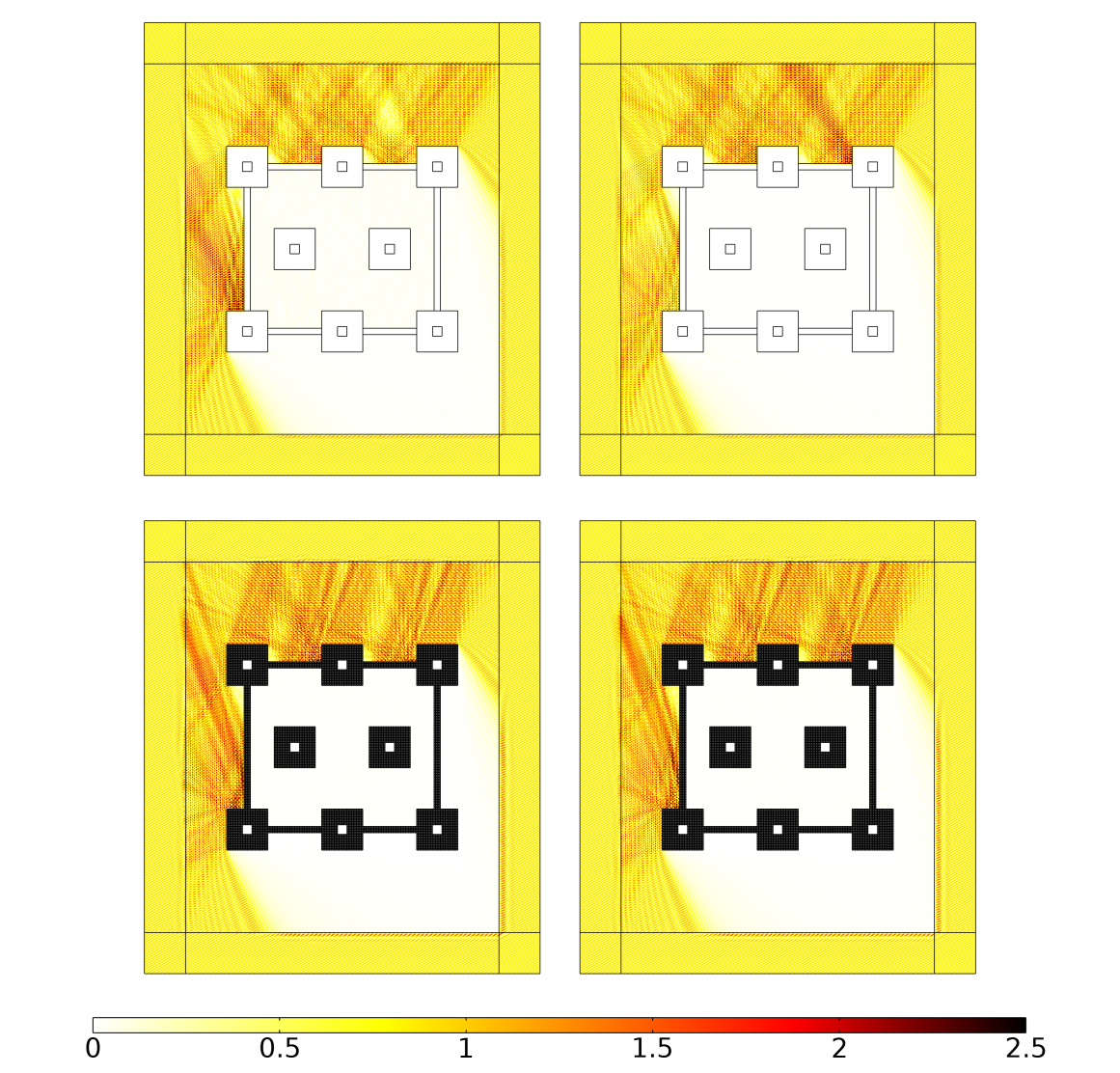}
\caption{Peripheral shield 4 unit cells thick for an incident wave with a frequency of $\omega=0.92$ MHz and an angle of $\phi=30^{\circ}$.	
Real (\textit{left}) and imaginary (\textit{right}) part of the normalized displacement for the 
(\textit{first row}) microstructured material and (\textit{second row}) equivalent micromorphic continuum. The thickness of each single shields is 10 unit cells.}
\label{fig:multiple_shields}
\end{figure}

Thanks to the relaxed micromorphic model presented so far, with the associated well-posed boundary conditions, we are now able to explore large-scale optimization problems of this type.
The results of Sect. \ref{sec:rmm_3} and \ref{sec:rmm_4} show an additional region with no displacement on the opposite side of the boundary of the shield that is first impacted by the incident wave.
We now explore a multiple-shields structure in order to investigate the displacement field in the region situated between and behind a group of single shields.
Fig.~\ref{fig:multiple-3x3} shows the global reflection pattern of the multiple-shields systems.
It is apparent that the configurations chosen are effective for local shielding of the internal small squares, as well as for the global shielding between the simple shields, as far as a 0$^{\circ}$ incidence is considered.
For other angles of incidence the global shielding effect clearly loses its effectiveness.
To optimize the global shielding effect an additional ``metamaterial's ring" can be added to the configuration of Fig.~\ref{fig:multiple_shields} and the local and global reflection patterns can be explored for different angles of incidence.
Such explorations would not be possible, or very time consuming, with simulations with the full-microstructure.
The simulation of Fig.~\ref{fig:multiple_shields} of the relaxed micromorphic model needed \textit{35 min 8 s}, while that of the microstructured case required \textit{2 h 11 min 46 s}.
The computational gain clearly grows fast as soon as larger structures are investigated.

\section{Conclusions}
In this paper, we showed that the relaxed micromorphic model can be effectively used to describe wave propagation and scattering in complex structures combining metamaterials' and classical materials' bricks of finite size.
We successfully designed a double-shield device extending the frequency range for which the screening effect occurs, thanks to the use of unit cells with different size.
We also designed a multiple-shield structure allowing to protect both the internal part of each shield and the external region between the shields.
Thanks to the model’s simplification (homogenized model featuring few frequency-independent parameters), these designs were obtained at a fraction of the computational cost with respect to classical simulations based on Cauchy elasticity.
This gain in computational time will be primordial for further works where more and more complex structures will be investigated with the aim of targeting real engineering applications.
The results of the present paper open the way to more systematic explorations of engineering large-scale structures that can control elastic waves, thus creating favorable conditions for effective energy conversion and re-use.

%%%%%%%%%%%%%%%%%%%%%%%%%%%%%%%%%%%%%%%%%%%%%%%%%%%%%%%%%%%%%%%%%%%%%%%%%%%%%%%%
%%%%%%%%%%%%%%%%%%%%%%%%%%%%%%%%%%%%%%%%%%%%%%%%%%%%%%%%%%%%%%%%%%%%%%%%%%%%%%%%
{\scriptsize
	\paragraph{{\scriptsize Acknowledgements.}}
	Angela Madeo, Leonardo A. Perez Ramirez, and Gianluca Rizzi acknowledge support from the European Commission through the funding of the ERC Consolidator Grant META-LEGO, N$^\circ$ 101001759.
    The authors also gratefully acknowledge the computing time provided on the Linux HPC cluster at Technical University Dortmund (LiDO3), partially funded in the course of the Large-Scale Equipment Initiative by the German Research Foundation (DFG) as project 271512359.
}

%%%%%%%%%%%%%%%%%%%%%%%%%%%%%%%%%%%%%%%%%%%%%%%%%%%%%%%%%%%%%%%%%%%%%%%%%%%%%%%%
%%%%%%%%%%%%%%%%%%%%%%%%%%%%%%%%%%%%%%%%%%%%%%%%%%%%%%%%%%%%%%%%%%%%%%%%%%%%%%%%

%%%%%%%%%%%%%%%%%%%%%%%%%%%%%%%%%%%%%%%%%%%%%%%%%%%%%%%%%%%%%%%%%%%%%%%%%%%%%%%%
%%%%%%%%%%%%%%%%%%%%%%%%%%%%%%%%%%%%%%%%%%%%%%%%%%%%%%%%%%%%%%%%%%%%%%%%%%%%%%%%

\begingroup
\setstretch{1}
\setlength\bibitemsep{3pt}
\printbibliography
\endgroup

%%%%%%%%%%%%%%%%%%%%%%%%%%%%%%%%%%%%%%%%%%%%%%%%%%%%%%%%%%%%%%%%%%%%%%%%%%%%%%%%
%%%%%%%%%%%%%%%%%%%%%%%%%%%%%%%%%%%%%%%%%%%%%%%%%%%%%%%%%%%%%%%%%%%%%%%%%%%%%%%%
\end{document}